\documentclass[12pt, prd, showpacs]{revtex4}
\usepackage{amsmath}
\usepackage{amssymb}
\usepackage{graphicx}
\usepackage{bm}

\input{tcilatex}

\begin{document}

\title{Horizon closeness bounds for static black hole mimickers}
\author{ Sergey V. Sushkov$^{a,b}$}
\email{sergey_sushkov@mail.ru; sergey.sushkov@ksu.ru}
\affiliation{$^a$Department of General Relativity and Gravitation, Kazan State
University, Kremlevskaya St. 18, Kazan 420008, Russia}
\affiliation{$^b$Department of Mathematics, Tatar State University of Humanities and
Education, Tatarstan St. 2, Kazan 420021, Russia}
\author{Oleg B. Zaslavskii}
\email{ozaslav@kharkov.ua}
\affiliation{Astronomical Institute of Kharkov V. N. Karazin National University, 35
Sumskaya St., Kharkov, 61022, Ukraine}

\begin{abstract}
We consider the question whether a wormhole can be converted into a
non-extremal quasi-black black hole by continuous change of parameters. In
other words, we ask whether \textquotedblleft black\textquotedblright\
wormholes can exist as end points of families of static wormhole geometries.
The answer is negative since the corresponding limit is shown to be
singular. Similar conclusions are valid also for other types of black hole
mimickers such as gravastars and quasi-black holes without wormhole
behavior. Our treatment is model-independent and applies to any static
geometries without requirement of special symmetries. We also find an
asymptotic expression for the Kretschmann scalar for wormholes on the
threshold of horizon formation that can be used as an the bound on proximity
of the configuration to the would-be horizon. The derived bound is very weak
for astrophysical black holes but becomes relevant for microscopic ones. We
point out complementarity between ability of wormholes to mimic black holes
and their ability to be traversable ``in practice''.
\end{abstract}

\maketitle




Black holes and traversable wormholes are the brightest examples of object
predicted by general relativity. Usually, these entities are considered as
opposed to each other since a black hole has an event horizon whereas a
traversable wormhole does not have it by definition. Meanwhile, the relation
between both types of objects becomes non-trivial in the situation when the
horizon is \textquotedblleft almost\textquotedblright\ formed. Such an issue
became relevant in the astrophysical context where the general question was
posed \cite{abr} -- to which extent observational data which can be
considered as an evidence of black holes existence are unambiguous? May they
correspond to compact bodies which behave very similarly to black holes but
do not have a horizon? This question can also be important in what concerns
the fate of wormholes in the course of evolution of Universe. It is stated
in the recent work \cite{kf} that such wormholes can evolve to form black
holes.

Several types of black hole mimickers were suggested - such as gravastars 
\cite{grav} (including their dark energy variety \cite{lobo}) and
quasi-black holes (see \cite{lz1} and references therein). In the recent
paper \cite{ds} wormholes were added to this list that gave rise to some
``hybrids'' of quasi-black holes and wormholes. It was proposed to consider
wormholes with the throat very close to the would-be horizon as a good
mimicker of a black hole (\textquotedblleft black hole
foils\textquotedblright ) that looks almost indistinguishable from a true
black hole. However, it was pointed out in \cite{lz2} that in the limit when
the throat of a wormhole approaches the would-be horizon, the geometry
becomes singular. This poses the question whether this feature remains in
the general (not necessarily spherically-symmetric) static geometry or not.
Physical relevance of this question stems from several points.

By its own definition, the horizon of a full-fledged black hole represents a
regular (hyper)surface. Therefore, if the limit under discussion is singular
(we show below that this is indeed the case) it actually means that a black
hole mimicker cannot be converted into a black hole. However, it does not
mean that such mimickers are irrelevant as good approximation to black
holes. By contrary, we derive the bound on proximity of the throat to the
horizon and show that for astrophysical black holes such a bound is
satisfied with huge margin, so that mimickers still can serve as a real
alternative to real black holes. The situation, however, drastically changes
for microscopic black holes. The issue in question is also interesting from
the viewpoint of the cosmic censorship since the limiting state in which the
curvature invariants diverge combines the features of a naked singularity
and of a horizon.

Thus, it seems that one can approach the limiting state under consideration
as closely as one likes but this limit itself is unattainable. To some
extent, this resembles the situation with the third law of black hole
thermodynamics where one can deform the non-extremal black hole but cannot
convert it into an extremal one with the final number of steps in any
physical process.

The fact that the limit is singular was reached previously for spherically
symmetrical space-time and for two kinds of models \cite{lz2}. In the
present work we establish a general model-independent proof valid for any
static space-time. Our results apply also to other types of black hole
mimickers -- for example, gravastars. (In the extremal case regular
mimickers are possible \cite{lz2} but they have very limited astrophysical
significance.) We also derive a simple universal bound in terms of the lapse
function and the surface gravity of the ``would-be horizon''.

One could naively rely on topological arguments: as a formation of \ a black
hole implies the appearance of a trapped surface, one could expect that such
a surface cannot appear "from nothing" as a limiting point of continuous
changes. However, the key point consists in that the state under discussion
is not a (supposedly existing) black hole but a \textit{quasi-}black hole.
The qualitative difference between them (which was not taken into account in 
\cite{kf}) consists in that a configuration forming a black hole collapses
under the horizon where an essentially non-static so-called T-region is
present and where the formation of trapped surfaces occurs. Meanwhile, the
quasi-black hole state assumes, by definition \cite{lz1}, that a system does
not collapse at all, no matter how close to the gravitational radius its
surface be. As a result, the trapped surface is not expected to form at all.
Thus, topological argument is insufficient.

We choose another criterion. Our approach consists in that we check whether
the geometry of a wormhole metric on the threshold of formation of the
horizon (almost \textquotedblleft black wormhole\textquotedblright ) remains
regular in this limit. If the answer is negative, this means that a wormhole
cannot be converted to a full-fledged black hole. To this end, we implement
the following strategy: (i) write the asymptotic form of the metric near the
would-be horizon in the appropriate coordinates, (ii) check the finiteness
of the Kretschmann scalar, (iii) check the finiteness of the surface
stresses on the throat if they do not identically vanish there. In doing so,
it is convenient to use an elaborated formalism of 2+1+1 splitting of
space-time which in the black hole limit should reproduce known results \cite%
{vis}.

Consider the static metric of the general form which can be written in Gauss
coordinates as%
\begin{equation}
ds^{2}=-N^{2}dt^{2}+dl^{2}+\gamma _{ab}dx^{a}dx^{b},\quad a,b=1,2,
\end{equation}%
where $l$ is a proper distance and $\gamma _{ab}=\gamma _{ab}(l,x^{a})$. We
assume that $N=N(l,x^{a};\varepsilon )$ where $\varepsilon $ is a parameter.
For $\varepsilon >0$ the metric does not have a horizon according to a
meaning of the traversable wormholes. It means that for small but non-zero $%
\varepsilon $ the lapse function $N$ remains non-zero where we put $l=0$ on
the throat. At $\varepsilon =0$ the horizon appears at $l=0$: $%
N(0,x^{a};0)=0 $. (We consider here non-extremal black holes, so that the
proper distance from the horizon is finite, so that we always can put $l=0$
on the horizon.)

Let us consider two branches of a wormhole (with increasing and decreasing
areas as functions of $l$ - in accordance with the definition of a throat in
generic static space-time \cite{dv}) glued at some two-dimensional surface
near the throat. This is a kind of the surgery approach (see, in particular,
Sec. 15. 2.1 of \cite{visbook}). In general, it generates the delta-like
shell with surface stresses located at the position of gluing. We consider
two situations. To retain succession with \cite{lz2}, we call them BT and TB
(capitals borrow here the first letters from ``black hole'' and ``throat'').
Situations BT: first we move the shell to the throat, then we let the
horizon to appear. Formally, it corresponds to the limiting transition $%
\lim_{\varepsilon \rightarrow 0}\lim_{l\rightarrow 0}$. Situation TB: first
we let the horizon to appear, then we move the position of the shell to the
horizon. Formally, it corresponds to the limiting transition $%
\lim_{l\rightarrow 0}\lim_{\varepsilon \rightarrow 0}$. (The simplest
example is gluing between two copies of the Schwarzschild metric \cite%
{visbook}). The key point consists in that both limits in general do not
coincide (see \cite{lz2} and below).

We are interested in the case when the lapse function $N>0$ for all $l$ and $%
\lim_{\varepsilon \rightarrow 0}N(0,x^{a},\varepsilon )=0$. The latter just
means that the configuration is on threshold of forming a horizon at the
location of the throat. Mathematically, the both aforementioned conditions
entail that for sufficiently small $\varepsilon $ the lapse function should
have a minimum with respect to $l$ since $N\rightarrow 0$ at $l=0$ and $N>0$
for $l\neq 0$ in the vicinity of $l=0$.

Correspondingly, we deal with two cases. 1) The minimum of the lapse
function as a function of $l$ is regular. 2) The minimum of the lapse
function is non-regular, its derivative with respect to $l$ has a jump.
Thus, we should consider 4 combinations: BT1, BT2, TB1, TB2. In Ref. \cite%
{lz2} this was done for the spherically symmetric space-times and for two
kinds of models only. Now, we show how this can be done in a
model-independent way and without requirement of spherical symmetry. Our
main point consists in showing that the limit in all of four cases becomes
singular. More precisely, we demonstrate that either the Kretschmann scalar
or the magnitude of the surface stresses become infinite.

The Kretschmann scalar $Kr=R_{\alpha \beta \gamma \delta }R^{\alpha \beta
\gamma \delta }$ is equal to (see \cite{vis}) 
\begin{equation}
Kr=P_{ijkl}P^{ijkl}+4\frac{N_{;ij}N^{;ij}}{N^{2}}  \label{kr}
\end{equation}%
where the curvature tensor $P_{ijkl}$ and covariant derivatives (...)$_{;i}$
refer to the spatial metric $g_{ij}$ of the hypersurface $t=const$. Now we
apply this requirement to the throat of the wormhole which, by assumption,
is approaching the horizon in the limit $\varepsilon \rightarrow 0$. As the
sum in (\ref{kr}) is taken with respect to positively defined spatial
metric, it is positively definite\textit{\ }and\textit{\ }the finiteness of $%
Kr$ is possible if and only if each term is bounded. In particular, it
entails that $C_{ij}=\frac{N_{;ij}}{N}$ should be finite.

Considering, for definiteness, the part of space-time with $l\geq 0$, we
must have 
\begin{equation}
\lim_{\varepsilon \rightarrow 0}N(l,x^{a};\varepsilon )=N_{BH}(l,x^{a})
\end{equation}%
where $N_{BH}(l,x^{a})$ corresponds to a black hole metric. Then, for $%
l\rightarrow 0$ the following expansion holds \cite{vis}:%
\begin{equation}
N_{BH}=\kappa _{+}l+\frac{N_{3}(x^{a})}{3!}l^{3}+...  \label{bh}
\end{equation}%
where $\kappa _{+}=const$ is a surface gravity. If we want to have a
wormhole, the similar expression holds for $l<0$ with $\kappa $ replaced by $%
-\kappa _{-}$ where $\kappa _{-}>0$.

For what follows we also need to have the expression for surface stresses
which are potentially present on the throat. According to general rules \cite%
{isr},

\begin{equation}
8\pi S_{0}^{0}=-[\![K_{a}^{a}]\!]\text{,}
\end{equation}%
\begin{equation}
8\pi S_{a}^{b}=[\![K_{a}^{b}]\!]-\delta _{a}^{b}[\![K_{0}^{0}+K_{c}^{c}]\!]
\label{sab}
\end{equation}%
where $[\![...]\!]\equiv (...)_{l=+0}-(...)_{l=-0}$, $K_{_{\mu \nu
}}=-\nabla _{\nu }n_{\mu }$ is the tensor of the extrinsic curvature, $%
n_{\mu }$ is the unit normal vector to the hypersurface $l=0$. The
finiteness of $Kr$ requires the finiteness of $C_{ab}$ that leads to the
fact that $K_{ab}\rightarrow 0$ in the horizon limit \cite{vis}. Therefore,
the only potential source of divergences is contained in the term\thinspace 
\begin{equation}
-[\![K_{0}^{0}]\!]=\frac{2}{N}\left[ \left( \frac{\partial N}{\partial l}%
\right) _{+}-\left( \frac{\partial N}{\partial l}\right) _{-}\right] \,\text{%
.}  \label{k}
\end{equation}

Case 1. It means that we have the Taylor expansion of the type which is more
convenient to write in terms of $-g_{tt}=N^{2}$ as 
\begin{equation}
N^{2}=N_{0}^{2}(x^{a};\varepsilon )+A_{2}(x^{a};\varepsilon
)l^{2}+A_{3}(x^{a};\varepsilon )l^{3}+...  \label{N2}
\end{equation}%
where all coefficients $A_{k}$ are well-defined in the limit $\varepsilon
\rightarrow 0$.

By comparison to (\ref{bh}),%
\begin{equation}
\lim_{\varepsilon \rightarrow 0}N_{0}^{2}(x^{a};\varepsilon )=0\text{, }%
\lim_{\varepsilon \rightarrow 0}A_{2}(x^{a};\varepsilon )=\kappa _{+}^{2}.
\label{lim}
\end{equation}

Then, simple calculations give us%
\begin{equation}
C_{ll}=\frac{N_{0}^{2}A_{2}+3A_{3}N_{0}^{2}l+O(l^{3})}{%
[N_{0}^{2}+A_{2}l^{2}+O(l^{3})]^{2}}\text{.}  \label{Cll}
\end{equation}

Situation BT. It follows from (\ref{lim}) and (\ref{Cll}) that 
\begin{equation}
\lim_{l\rightarrow 0}C_{ll}=\frac{A_{2}(x^{a};\varepsilon )}{%
N_{0}^{2}(x^{a};\varepsilon )}\text{.}  \label{c11}
\end{equation}

Then, we obtain that and $\lim_{\varepsilon \rightarrow 0}\lim_{l\rightarrow
0}C_{ll}=\infty $. As a result, $Kr$ also diverges and thus, the
configuration becomes singular.

Situation TB. Then, $\lim_{l\rightarrow 0}\lim_{\varepsilon \rightarrow
0}C_{ll}$ is finite as well as the quantity $Kr$. However, now we will see
that there are infinite surface stresses. Indeed, in this situation $%
N=\kappa _{+}l+O(l^{3})$ for $l>0$ and $N=-\kappa _{-}l+O(l^{3})$ for $l<0$.
As a result, it follows from (\ref{sab}), (\ref{k}) that the quantity $%
S_{a}^{b}\rightarrow \infty $.

Case 2. The minimum of the lapse function is non-regular, its derivative
with respect to $l$ has a jump. Then,%
\begin{equation}
N=N_{0}(x^{a};\varepsilon )+N_{1}^{+}(x^{a};\varepsilon )l+...  \label{+}
\end{equation}%
for $l>0$ and%
\begin{equation}
N=N_{0}(x^{a};\varepsilon )+N_{1}^{-}(x^{a};\varepsilon )l+...  \label{-}
\end{equation}%
for $l<0$. Here the absence of a horizon entails 
\begin{equation}
N_{1}^{+}(x^{a};\varepsilon )>0\text{, }N_{1}^{-}(x^{a};\varepsilon )<0.
\label{+-}
\end{equation}

In the limit $\varepsilon \rightarrow 0$ following the same lines as in \cite%
{vis} and considering the behavior of $c_{ab}$ we obtain that $%
\lim_{\varepsilon \rightarrow 0}N_{1}^{\pm }=\pm \kappa ^{\pm }$ where $%
\kappa ^{\pm }>0$ are constants.

In this case the evaluation of ${K\!r}$ gives no useful information since it
can be finite in the limiting transition under discussion. However, the
singular behavior reveals itself in surface stresses.

In our case $\left( \frac{\partial N}{\partial n}\right) _{+}=N_{1}^{+}\neq
N_{1}^{-}=\left( \frac{\partial N}{\partial n}\right) _{-}$ and even have
different signs, so that $\left( \frac{\partial N}{\partial n}\right)
_{+}-\left( \frac{\partial N}{\partial n}\right) _{-}\rightarrow \kappa
_{+}+\kappa _{-}>0$ in the limit $\varepsilon \rightarrow 0$. But the
denominator in (\ref{k}) tends to zero in this limit. As a result, the
stresses $S_{a}^{b}\rightarrow \infty $ both in situations BT and TB.

It is worth noting that in our approach we did not use the geometrical
properties of the throat itself (see \cite{dv} for definition and discussion
of generic static wormholes). What is important in our treatment is the
behavior of the lapse function of the ``almost black hole'' with $%
\varepsilon \neq 0$ that ensures the absence of the horizon. For this
reason, the same conclusions apply to other types of black hole mimickers.
This generalizes previous observations made in \cite{lz2}, \cite{lz1} for
spherically-symmetrical quasi-black holes and gravastars \cite{grav} where
the role of the unbound stresses near the would-be horizon was pointed out.
(More precisely, if we instead of a wormhole metric assume the usual
monotonic dependence of the area on the proper distance, the behavior of the
lapse function for quasi-black holes is different in that $N\rightarrow 0$
everywhere in the inner region \cite{lz1} but this leads to unbound stresses
also.) The case of wormholes is more interesting in that even in the absence
of thin shells the geometry turns out to be singular in the horizon limit
anyway. This singularity occurs in the limit of succession of families of
time-like surfaces accessible to an outer observer (although the limit
itself realizes on the light-like surface). In this sense, the impossibility
of the existence of legitimate black hole mimickers is related to the
principle of the cosmic censorship.

Thus, we gave a general model-independent proof concerning impossibility to
convert a regular mimicker to a full-fledged static non-extremal black hole.
However, nothing prevents the existence of extremal black holes obtained by
the limiting transitions from wormholes. For spherically-symmetrical
configurations, concrete model examples were suggested in \cite{lz2}. Such
configurations can also occur in the context of non-minimal
Einstein-Yang-Mills theories \cite{bsz}. But, for relativistic astrophysics,
black holes are considered usually to be non-extremal.

Our results do not exclude the relevance of black hole mimickers from a
practical viewpoint in some approximation (when a surface is close to the
horizon with some fixed precision). Here, our results entail the universal
asymptotic behavior of the Kretschmann scalar and tidal forces. Namely,
consider for definiteness case 1, situation BT which is more interesting
from the astrophysical viewpoint since the divergences are now due of the
curvature invariant and not because of the stresses on the shell. Then, in
the limit under discussion $\varepsilon \rightarrow 0$, $N_{0}\rightarrow 0$
it follows from (\ref{kr}), (\ref{lim}), (\ref{c11}) that%
\begin{equation}
\sqrt{Kr}\approx \frac{2\kappa ^{2}}{N_{0}^{2}}  \label{limkr}
\end{equation}%
where we omitted the subscript at $\kappa $ that has a meaning of the
surface gravity of the would-be black hole towards which the configuration
approaches. In doing so, it is the longitudinal tidal forces (in the $l$
direction) that give the main contribution. Further application depends on
details of concrete non-gravitational physics. Let $\sqrt{Kr}<p_{0}=\frac{1}{%
L^{2}}$ where $L$ is a typical scale factor in the sense that either the
material of the system cannot withstand larger gravitational forces on small
distances or there physics is simply unknown (say, the role of $L$ is played
by the Planck length; in the latter case the backreaction of quantum fields
may affect the geometry significantly), $p_{0}$ being the corresponding
effective pressure. Then, we obtain the limitation on the proximity of the
geometry to the horizon (the minimum value of the lapse function): $N_{0}>%
\sqrt{2}\kappa L$.

For simplicity, let us consider the spherically-symmetrical case. Then, one
can obtain easily that the Kretschmann scalar is dominated by the transverse
pressure, $\kappa \sim M^{-1}$ ($M$ is a mass). Now, $N_{0}>\frac{L}{M}$ (in
geometrical units with fundamental constants $G=c=1$). For estimations we
may take $M$ to be typical of astrophysical black holes, say $\sim
M_{\bigodot}\approx 10^5cm$. As for the characteristic length of the
gravitational field, we assume that $L$ is of order of typical atomic sizes,
i.e. $\sim 10^{-8}cm$. On these scales tidal forces would destroy atomic
structure of matter, and, as a consequence, a spectroscopic appearance of
such astrophysical objects would be drastically changed. Estimating now
yields $N_0>10^{-13}$; this bound gives no useful restriction on analysis of
observational data. 
%
However, if $M\sim m_{Pl}$ (the Planck mass) and $L\sim l_{Pl}$ (Planck
length), this bound becomes significant. In particular, it concerns dynamic
scenarios of formation of alternatives to black holes where quantum
backreaction is essential \cite{bar}. If we summarize the consequences of
the bound we can see that (i) it confirms that alternative to black holes
remain viable candidates from the practical observational viewpoint, (ii)
the situation can drastically change, say, for early universe where
microscopic black holes can be important, (iii) the result of general
character consists in the impossibility to convert a wormhole to a
full-fledged black hole.

We would also like to draw attention that there is some complementarity
between the ability of a wormhole to mimic a black hole and its property to
remain traversable "in practice". This is due to the factor $N_{0}^{-2}$
that enhances the tidal forces significantly as compared to, say, zero-tidal
forces solutions \cite{mt}. Therefore, the better a wormhole mimics a black
hole, the worse its traversability becomes. As a result, a traversable
wormhole can look as almost a black hole for an external observer but not as
a full-fledged wormhole for a free-falling observer.

In the present work we restricted ourselves by a pure static situation. The
separate issues which need separate treatment is the inclusion of rotation
and dynamics into consideration. What is more important, the full analysis
implies consideration of dynamic scenarios in the course of which wormhole
can (or cannot) evolve toward a quasi-black hole \cite{bar} state.

\begin{acknowledgments}
\vskip6pt S.S. was funded by the Russian Foundation for Basic Research
through the grant No 08-02-00325.
\end{acknowledgments}

\end{document}